\documentclass[fleqn,twoside]{article}
\usepackage{espcrc2}

\usepackage{graphicx}

\title{Heavy Quarkonia Results from CLEO}

\author{Kamal K. Seth\address[MCSD]{Department of Physics and Astronomy, 
        Northwestern University\\Evanston, IL, 60208, USA}}
       
\begin{document}

\begin{abstract}
The latest experimental results in bottomonium and charmonium spectroscopy from CLEO is presented.
\end{abstract}

\maketitle

Heavy quarkonium $c\bar{c}$ ($c\bar{c}$ charmonium, $b\bar{b}$ bottomonium) provides the best means of testing QCD, both the validity of perturbative QCD and potential models, and lattice QCD calculations.  Bottomonium is better than charmonium, both because it has smaller relativistic problems\hspace{0.8cm} ($<v^2/c^2>\approx0.1$ versus $\approx 0.2$) and smaller coupling constant ($\alpha\approx0.2$ versus $\approx0.35$), but much less high precision spectroscopic information is available for bottomonium.  No $b\bar{b}$ singlet states are known, and very few hadronic and radiative decays are known.  Nevertheless, progress is being made through recently taken $\Upsilon(nS)$ data with CLEO III with much larger luminosity than before, the event counts being $\Upsilon(1S)\sim20$ million, $\Upsilon(2S)\sim10$ million, and $\Upsilon(3S)\sim50$ million.  With the beginning of the CLEO-c program, interesting new results are also being produced in the charmonium region, $\psi'(2S)\sim$ 3 million.

I will now review some of the most recent results, first in bottomonium spectroscopy, and then in charmonium spectroscopy.

\section{First Observation of New $\Upsilon(1D)$ State of Bottomonium}

No D-wave ($l=2$) states have ever before been identified in bottomonium, though both $1D_J$ and $2D_J$ states are expected to be bound.  CLEO has now observed the $\Upsilon(1D)$ state in the following four photon cascade[1]:
\begin{eqnarray*}
\Upsilon(3S)\to\gamma_1\chi(2P)\to\gamma_1\gamma_2\Upsilon(1D)\to\gamma_1\gamma_2\gamma_3\chi(1P)\\\to\gamma_1\gamma_2\gamma_3\gamma_4\Upsilon(1S)\to\gamma_1\gamma_2\gamma_3\gamma_4(e^+e^-,\mu^+\mu^-),
\end{eqnarray*}
which is illustrated in Fig. 1.  The number of signal events observed was $34.5\pm6.4$ and the significance of the signal was $10.2\sigma$. \hspace{1in} M[$\Upsilon(1^3$D$_2$) =  $10161.1\pm0.6\pm1.6$ MeV, and  $B(\gamma\gamma\gamma\gamma l^+l^-)_{\Upsilon(1D)} = (2.6\pm0.5\pm0.5\pm) \times 10^{-5}$. 
\begin{figure}
\includegraphics[width=2.5in]{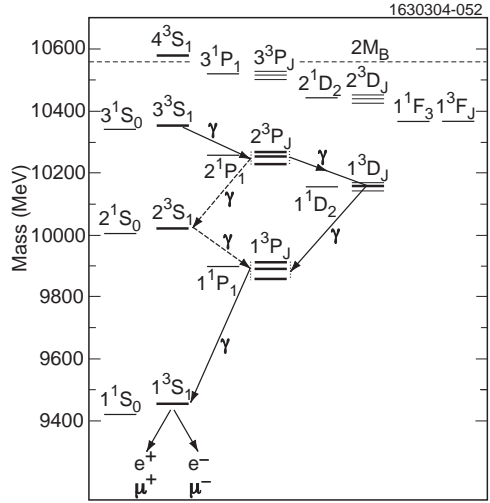}
\vspace*{-0.65cm}
\caption{$1^3D_2$ detection in 4-photon cascade.}
\vspace*{-0.65cm}
\end{figure}

\section{Measurement of $B[\Upsilon(nS)\to\mu^+\mu^-]$}

The important parameters of $\Upsilon(2S)$ and $\Upsilon(3S)$ resonances, that is, leptonic widths, $\Gamma_{ee}$, and total widths, $\Gamma$, are not well established.  Measurement of $\Upsilon(nS)$ decay to muon pairs relative to hadrons near resonance peaks gives
$$\tilde{B}_{\mu\mu} = \frac{\Gamma_{\mu\mu}}{\Gamma_{had}} = \frac{N/\Upsilon\to\mu^+\mu^-)/\epsilon_{\mu\mu}}{N/\Upsilon\to hadrons)/\epsilon_{had}}$$
Assuming lepton universiality, 
$$B_{\mu\mu} = \Gamma_{\mu\mu}/\Gamma = \tilde{B}_{\mu\mu}/1+\tilde{B}_{\mu\mu}.$$  

$B_{\mu\mu}[\Upsilon(1S,2S,3S)]$ have been measured using this method [2].  The results are shown in the table below.  We note that while the $\Upsilon(1S)$ result is in good agreement with the PDG04 [3] value, the new results from $\Upsilon(2S)$ and $\Upsilon(3S)$ have much higher precision, and are quite different than the PDG04 values.  We also note that analysis of CLEO III resonance scans (in progress) will provide separate measurements of $\Gamma_{ee}$ and therefore lead to precision measurements of $\Gamma=\Gamma_{ee}/B_{\mu\mu}$.

\begin{tabular}{|r|c|c|}
\hline
 & $B_{\mu\mu}(\%)$ CLEO & $B_{\mu\mu}(\%)$ PDG\\
\hline
$\Upsilon(1S)$ & $2.49\pm0.02\pm0.07$ & $2.48\pm0.06$ \\
$\Upsilon(2S)$ & $2.03\pm0.03\pm0.08$ & $1.31\pm0.21$ \\
$\Upsilon(3S)$ & $2.39\pm0.07\pm0.10$ & $1.81\pm0.17$ \\
\hline
\end{tabular}

\section{$\Upsilon(1S)$ decays to Charmonia}

$\Upsilon(1S)$ decays to $J/\psi$, $\psi'$, and $\chi_{c1,c2}$ have been measured.  The decay $\Upsilon(1S)\to J/\psi+X,\;J/\psi\to e^+e^-,\mu^+\mu^-$ leads to [4]
$$B[\Upsilon(1S)\to J/\psi+X] = (6.4\pm0.4\pm0.6)\times10^{-4},$$
which is nearly a factor two smaller than the current PDG04 value of $(11\pm4)\times10^{-4}$, and has much smaller errors.

The $J/\psi$ momentum distribution is found to be in clear disagreement with the prediction based on the color octet model [5], and in qualitative agreement with the color singlet model.

\section{First Observation of $\chi_b'(2P)\to \omega\Upsilon(1S)$}

For the first time in a bottomonium system, a hadronic transition other than $\Upsilon(nS)\to\Upsilon(n'S)\pi\pi$ has been observed [6], with
\begin{eqnarray*}
B[\chi_{b1}(2P)\to\omega\Upsilon(1S)] = (1.68\pm0.38)\%\\
B[\chi_{b2}(2P)\to\omega\Upsilon(1S)] = (1.10\pm0.34)\%
\end{eqnarray*}


\begin{figure}
\includegraphics[width=2.5in]{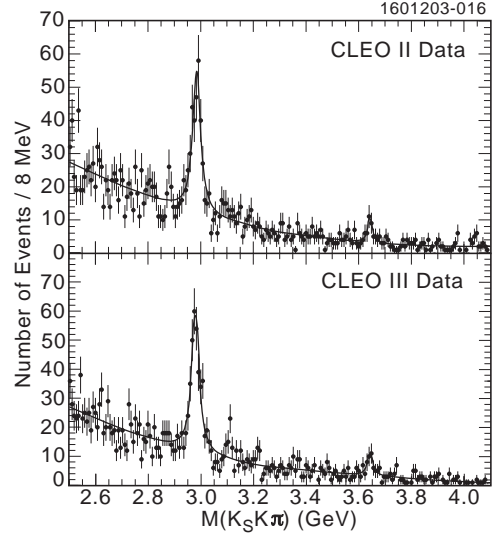}
\vspace*{-0.65cm}
\caption{$K_S K \pi$ invariant mass distributions for CLEO II and CLEO III independent data sets.}
\vspace*{-0.65cm}
\end{figure}

\section{Discovery of $\eta_c'(2S)$}

The Belle experiment observed $\eta_c'(2S)$ in two different channels with a mass average of [7]:
$$M(\eta_c')=3641\pm8\;\mathrm{MeV}$$

CLEO [8] and BaBar [9] have observed $\eta_c'(2S)$ in two photon fusion
$$e^+e^-\to(e^+e^-)\gamma\gamma;\quad \gamma\gamma\to(\eta_c,\;\eta_c')\to K_S K^\pm \pi^\mp$$
The $K_S K \pi$ missing mass distributions for the CLEO measurements are shown in Fig. 2.

CLEO [8] measures the parameters $M=3642.9\pm3.1\pm1.$5 MeV, $\Gamma < 31$ MeV (90\% CL) and $\Gamma_{\gamma\gamma} = 1.3\pm0.6$ keV.  

The average of Belle, BaBar, and CLEO measurements is $M[\eta_c'(2S)]=3637.4\pm4.4$ MeV.  

The $2S$ hyperfine mass splitting is therefore,
\vspace*{-6pt}
$$\Delta M(2S) = M[\psi'] - M[\eta_c'] = 48.6\pm4.4 \; \mathrm{MeV}$$
We note that,\vspace*{-6pt}
$$\Delta M(1S) = M[J/\psi] - M[\eta_c] = 117\pm2 \mathrm{MeV}$$
The measured $\Delta M(2S)$ is thus much smaller than $\Delta M(2S)$, and also from most of the theoretical predictions.  It should lead to new insight into coupled channel effects on masses, and on the spin-spin contribution of the confinement part of the $q\bar{q}$ potential.

\section{Two-body Hadronic Decays of $\psi'(2S)$}

The pQCD expectation is that [assuming $\alpha_s(\psi'(2S))=\alpha_s(J/\psi)$] the ratio of light hadron decays of $\psi'(2S)$ and $J/\psi(1S)$ is
$$Q_{LH} \equiv \frac{B[\psi'\to LH]}{B[J/\psi\to LH]} \approx \frac{B[\psi'\to e^+e^-]}{B[J/\psi\to e^+e^-]} \approx 13\%$$
As a matter of fact, $Q(\Sigma H)_{expt} \approx (17\pm3)\%$.  However,for individual hadronic decays this ``13\%'' rule is found to be badly broken ($Q_{LH} \approx 0.2-20\%$).  No simple pattern is identifiable in the breaking of this rule in terms of the nature of the final state light hadrons.

BES has measured a large number of hadronic decays.  CLEO has now added many more [10], and found many more examples of strong violation of this rule.  Many possible theoretical explanations of the variations have been offered, but there is no consensus.  Perhaps, like the strong violations of the Hadron Helicity Conservation rule of QCD which have been observed in the $p\bar{p}$ formation and decay of spin 0 states $\eta_c$ and $\chi_0$, this is another example of the inapplicability of pQCD in three gluon annihilations.

\vspace*{-4pt}

\section{Radiative Transitions from $\psi'(2S)$}

CLEO has measured the inclusive photon spectrum from $\psi'(2S)$ decay [11].  The $B(\psi'(2S)\to\gamma\chi_{cJ}(1P))$ results are $(9.33\pm0.14\pm0.61)\%$, $(9.07\pm0.11\pm0.54)\%$, and $(9.22\pm0.11\pm0.46)\%$ for $J=2$, 1 and 0 respectively.  They are significantly higher than the values obtained by the PDG by a global fit to the $\psi'(2S)$ data [3], but agree well the previous measurements by the Crystal Ball [12], and have improved statistical and systematic errors.  We also observe the $M1$ transition to $\eta_c$, and obtain $B(\psi'(2S)\to\gamma\eta_c)=(0.32\pm0.04\pm0.06)\%$, but we do not observe the hindered $M1$ transition to $\eta_c'$ claimed by the Crystal Ball [12].
\vspace*{-4pt}

\section{Radiative Cascade Decays, $\psi'\to\gamma\gamma J/\psi$}

CLEO-c has accumulated $\sim6$ pb$^{-1}$ of data at $\psi'$. With $\sim$3 million $\psi'$ decays, precision determination of branching ratios for the decay channels $B(\chi_{cJ}\to\gamma J/\psi)$, $B(\psi'\to\eta J/\psi)$, $B(\psi'\to\pi^0 J/\psi)$ are being made from two photon cascades. 
\vspace*{-4pt}

\section{New Narrow State $X(3872)$}
\begin{figure}
\includegraphics[width=2.5in]{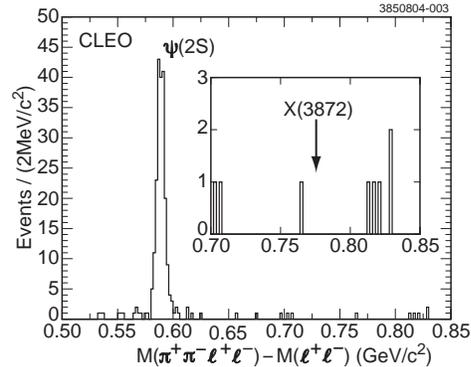}
\vspace*{-0.65cm}
\caption{Data events as function of $\Delta M\equiv M(\pi^+\pi^-l^+l^-) - M(l^+l^-)$.  The $\psi'(2S)$ is clearly visible and no apparent ``enhancement'' is seen in the $X(3872)$ region.}
\vspace*{-0.5cm}
\end{figure}

The Belle Collaboration has recently observed a narrow state $X(3872)$ decaying into $J/\psi + \pi^+\pi^-$ [13].  This observation has been confirmed by Babar, D0 and CDF [14-16].  All measurements are consistent with
$$M(X)=3872\pm1\;\mathrm{MeV}\;,\;\Gamma(X)\le2.3\;\mathrm{MeV}$$
Many proposals for the nature of this state have been made
\begin{itemize}
\setlength{\itemsep}{0pt}
\item a conventional charmonium state? [17]
\item a $D^0-\bar{D}^{*0}$ molecule? [18]
\item an exotic state, hybrid, or glueball?
\end{itemize}
It is important to identify the quantum numbers of $X(3872)$ in order to understand its nature.

CLEO has searched for $X(3872)$ state in untagged $\gamma\gamma$ fusion (+C parity, $J^{PC} = 0^{++}, 0^{-+}, 2^{++}, 2^{-+}, ...$), and in ISR production \hspace{0.4cm} (\quad $J^{PC} = 1^{--}$) with $\sim 15$ fb$^{-1}$ of CLEO III data [19]. Exclusive channels $X\to\pi^+\pi^-J/\psi, \; J/\psi\to l^+l^-$ were analyzed (see Fig. 3).

No signal of $X(3872)$ was found and the following 90\% confidence upper limits were set:

\noindent Untagged $\gamma\gamma$ fusion:
$$(2J+1)\Gamma_{\gamma\gamma}B(X\to\pi^+\pi^-J/\psi)<12.9 \, \mathrm{eV}$$
ISR production ($J^{PC}=1^{--}$):
$$\Gamma_{ee}B(X\to\pi^+\pi^-J/\psi)< 8.3 \, \mathrm{eV}$$

\section{Summary}

Heavy quarkonium spectrocopy continues to produce precision results.  With CLEO-c, we can look forward to the observation of charmonium $^1P_1(h_c)$ state, the search for charmonium $2P$ and $1D$ states, improved understanding of $D$ decays of $\psi''(3770)$ and $\psi'''(4040)$, and glueballs and other exotics.


\end{document}